# Effects of off-diagonal permittivity terms on polarization singularities in anisotropic grating system

Siyu Lei[1], Ze-Huan Zheng[2], Qilin Duan[3], Feng Wu[4], Xin Gao[1], Huanyang Chen[3], Ying Chen[1,*]

[1] Department of Physics, College of Information Science and Engineering, Huaqiao University, Xiamen 361021, China
[2] Institute of Electromagnetics and Acoustics, Xiamen University, Xiamen 361005, China
[3] Department of Physics, College of Physical Science and Technology, Xiamen University, Xiamen 361005, China
[4] School of Optoelectronics and Engineering, Guangdong Polytechnic Normal University, Guangzhou 510665, China

*Email: yingchen@hqu.edu.cn

**Abstract**

The evolutions of polarization singularities, including bound states in the continuum (BICs) and circularly polarized states (C points), are usually realized by tuning the geometric parameters of photonic crystal slabs. Here, we use the off-diagonal terms of permittivity tensor to manipulate polarization singularities without breaking the structural symmetry in an anisotropic grating system. By controlling the optical axis of anisotropic media, BICs can be shifted to different positions or split into C points, meanwhile, the creation and annihilation of multiple C points are also observed during the evolution process for both TE and TM modes, respectively. Remarkably, two different splitting directions of BICs can be achieved by tuning the off-diagonal terms of permittivity tensor for the two modes. This work illustrates the important role of off-diagonal terms on the far-field polarization singularities and provide an alternative way to precisely manipulate optical singularities.

## 1. Introduction

Off-diagonal anisotropic materials [1,2], including the off-diagonal permittivity tensor and permeability tensor, provide an important avenue for the manipulation of band dispersions, near- and far-field distributions, such as the topological bandgaps [3], Dirac and exceptional points [4,5], shear phonon polaritons [6,7] and topological polarization singularities [8-10]. There are several ways to achieve the off-diagonal anisotropic parameters, such as in natural materials, β-$Ga_2O_3$ inherently has off-diagonal terms of permittivity tensor that cannot be eliminated by coordinate transformations [11], while in magneto-optical materials, the gyromagnetic anisotropy of permeability tensor can be obtained by applying external magnetic field [3]. In the above applications of anisotropic materials, the polarization singularities with nonzero topological charges at the center have garnered great interest of many researchers and become the focus of our attention.

Recently, various polarization singularities in photonic crystal slabs (PCS), including bound states in the continuum (BICs), circularly polarized states (C points) and degenerate states, are widely studied due to their potential applications in light-trapping, chiral emission and vortex beam generation [12-24]. For the BICs with integer topological charges (V points), they are vortex centers of polarization vector for the far-field radiation, which cannot be radiated out and own the high quality factors [12, 13]. It has been demonstrated that the symmetry breaking usually causes the BICs to split into paired C points, and they obey the law of topological charge conservation [25-33]. For the C points with half-integer topological charges, their rich evolutions with structural and material parameters also reveal many interesting physical phenomena in momentum space [27, 30-33]. Specifically, it is worth mentioning that anisotropic materials play an important role in the study of polarization singularities. For example, in the magneto-optical PCS, the arbitrarily polarized BICs [10] and spin-orbit-locking chiral BICs [34] can be achieved by applying external magnetic field to break the time-reversal symmetry of PCS, thus realizing the magnetic modulation of BICs. In anisotropic waveguiding structures, various BICs and degenerate states are observed in the dispersion bands at different anisotropic parameters [4,5,35,36], meanwhile, in anisotropic cavities and PCS, topological half vortices can also appear by adjusting the optical axis (OA) of the anisotropic medium [37, 38]. However, the effects of different off-diagonal terms of permittivity tensor on various polarization singularities have not been explored and comparatively analyzed, which is a new degree of freedom and worth exploring new strategies to manipulate polarization singularities.

In this paper, we design an off-diagonal anisotropic grating lattice and investigate the evolution behaviors of three BICs (one symmetry-protected BIC and two accidental BICs) and multiple C points under different anisotropy, where the off-diagonal permittivity tensors are controlled by rotating the optical axis (OA) of the uniaxial material. By tracing the positions of BICs and C points in momentum space, we elaborate on the effects of different off-diagonal terms of permittivity tensor on the splitting of BICs and merging of C points for both TE and TM modes, and observe a complete process of the generation, shift, annihilation and merging of topological charges. In detail, we find that BICs will split into pairs of C points when the out-of-plane anisotropy symmetry is broken ($\varepsilon_{yz} \neq 0$ for TE mode; $\varepsilon_{xz} \neq 0$ and $\varepsilon_{yz} \neq 0$ for TM mode), while there is no BIC splitting when only the in-plane anisotropy symmetry reduces (that is $\varepsilon_{xy} \neq 0$). Our study shows that the off-diagonal terms are an important degree of freedom to control the polarization singularities, which may bring new scheme to the dynamic modulation of chiral optical singularities.

## 2. Results and discussion

The proposed off-diagonal anisotropic grating lattice is shown in Fig. 1(a), which is periodic in the x direction and uniform in the y direction. Here, the period of unit cell is fixed at P=900 nm, the height of the grating system is d=1.1P and the filling ratio of material A is $\rho$=0.46P. The relative permittivities of the alternating materials A and B are denoted as $\tilde{\varepsilon}_A$ and $\varepsilon_B$, respectively, with the surrounding background taking as air. Here, we set the relative permittivity of material B as $\varepsilon_B$=4, and the initial relative permittivity tensor of material A is represented $\tilde{\varepsilon}$=diag($\varepsilon_1$, $\varepsilon_2$, $\varepsilon_1$) when the OA is oriented along $x$ direction, characterizing the material as a uniaxial crystal with $\varepsilon_1$=9 and $\varepsilon_2$=9.2. As indicated in the coordinate system, the OA orientation within the anisotropic material A is controlled by the polar angle α and azimuth angle β, and the permittivity tensor for an arbitrary OA orientation can be obtained through transforming $\tilde{\varepsilon}$ by the rotation matrices R(α) and R(β), which can be written as (see more details in Appendix A)

$$\tilde{\varepsilon}_A = \begin{bmatrix} \varepsilon_{xx} & \varepsilon_{xy} & \varepsilon_{xz} \\ \varepsilon_{yx} & \varepsilon_{yy} & \varepsilon_{yz} \\ \varepsilon_{zx} & \varepsilon_{zy} & \varepsilon_{zz} \end{bmatrix} = \begin{bmatrix} \varepsilon_1\cos(\alpha)^2 + \varepsilon_2\sin(\alpha)^2 & \Delta\sin(\alpha)\cos(\alpha)\cos(\beta) & -\Delta\sin(\alpha)\cos(\alpha)\sin(\beta) \\ \Delta\sin(\alpha)\cos(\alpha)\cos(\beta) & \Delta\cos(\alpha)^2\cos(\beta)^2 + \varepsilon_1 & -\Delta\sin(\beta)\cos(\beta)\cos(\alpha)^2 \\ -\Delta\sin(\alpha)\cos(\alpha)\sin(\beta) & -\Delta\sin(\beta)\cos(\beta)\cos(\alpha)^2 & \Delta\cos(\alpha)^2\sin(\beta)^2 + \varepsilon_1 \end{bmatrix}, \quad (1)$$

where $\Delta = \varepsilon_2 - \varepsilon_1$. Thus, the polar angle α and azimuth angle β may break the anisotropy symmetry in and out of the x-y plane, respectively, and the off-diagonal terms are thus introduced by coordinate transformations. For such anisotropic material A with relative permittivity as $\tilde{\varepsilon}_A = [\varepsilon_{xx}\ \varepsilon_{xy}\ \varepsilon_{xz};\ \varepsilon_{yx}\ \varepsilon_{yy}\ \varepsilon_{yz};\ \varepsilon_{zx}\ \varepsilon_{zy}\ \varepsilon_{zz}]$ and relative permeability $\mu_r$=1, the governing equation inside material A can be written as $k^2\mathbf{E} - k(k \cdot \mathbf{E}) - \omega^2\mu_0\varepsilon_0\tilde{\varepsilon}_A\mathbf{E} = 0$ [39], where ω is the angular frequency. From the eigensolutions, we can find that for TE mode with $E_y \neq 0$, $E_x = E_z = 0$, the eigenfields are dependent of $\varepsilon_{yy}$, $\varepsilon_{xy}$, $\varepsilon_{zy}$; while for TM mode with $E_y = 0$, $E_x = E_z \neq 0$, the eigenfields are related to all three off-diagonal terms $\varepsilon_{xy}$, $\varepsilon_{xz}$ and $\varepsilon_{yz}$ and two main-diagonal terms $\varepsilon_{xx}$ and $\varepsilon_{zz}$ (see more details in Appendix B).

### 2.1 Evolution behaviors of polarization singularities for TE mode

First, we study the TE mode and discuss the different roles of $\varepsilon_{yy}$, $\varepsilon_{xy}$ and $\varepsilon_{zy}$ on the modulation of far-field polarizations in such grating system. The calculated TE band dispersions with the material anisotropy as α = 0º and β = 0º are illustrated in Fig. 1(b). For this case, the off-diagonal terms of permittivity tensor are all zero. It is shown that three BICs with integral topological charges are situated in the blue band (i.e., $TE_1$ band), indicating the occurrence of one symmetry-protected BIC at Γ point and two accidental BICs at off-Γ points. The inset displays the electric field distributions for the eigenstate at Γ point in $TE_1$ band, which reveals the strong confinement within the PCS. Fig. 1(c) further gives the polarization map and Q factors for the TE₁ band in Fig. 1(b). It can be found that there are three vortex polarization singularities locate at the high symmetry line $k_y$=0 with their positions at ($k_x$, $k_y$)=(0, 0) 2π/P and (±0.042, 0)2π/P in the momentum space, respectively. As we know, the topological charge $q$ for polarization singularities can be expressed as [12]

$$q = \frac{1}{2\pi} \oint_L dk_\| \cdot \nabla_{k_\|} \varphi(k_\|), \qquad (2)$$

where $L$ is the closed loop around singularities in the momentum space, $\varphi$ is the orientation angle between the polarization major axis and x axis. From the polarization vectors in Fig. 1(c), it is obvious that the three polarization singularities are BICs with topological charge of ±1, and the infinite Q factors appear at the three positions also indicate existence of BICs. Moreover, the polarization-orientation-angle diagram in the inset further determine that the Γ-BIC has a +1 charge, while the off-Γ BICs have a -1 charge.

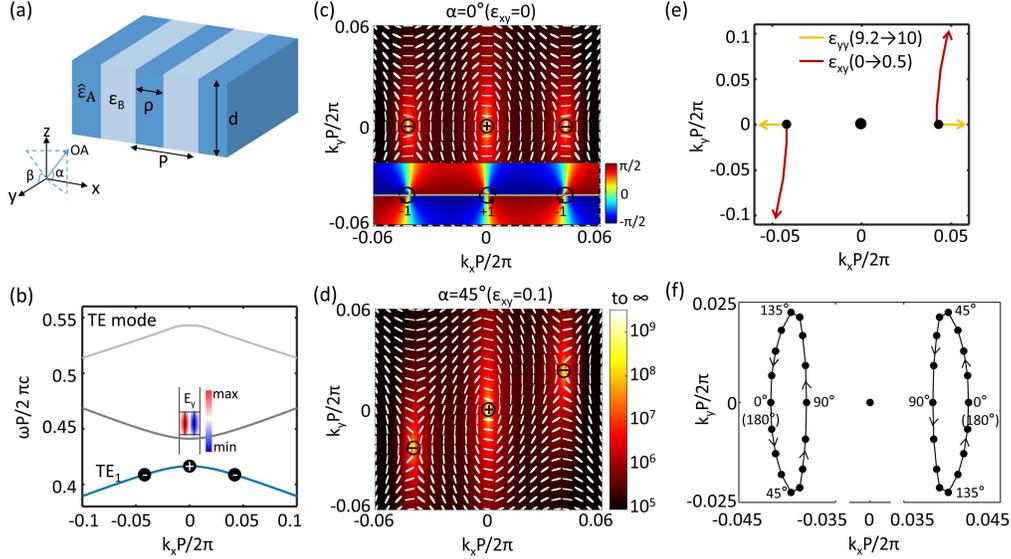

Fig. 1 (a) Schematic of an anisotropic grating lattice. The OA orientation (the blue arrow) for anisotropic material A is shown in the bottom left, with polar angle α and azimuth angle β in spherical coordinates. (b) The simulated TE band structures at α=0° and β=0° of the anisotropic grating lattice. The inset shows the distribution of $E_y$ field components for the $TE_1$ band at Γ point. (c) Calculated polarization map with Q factor as background for $TE_1$ band at α=0° and β=0°, and polarization-orientation-angle diagram around the three BICs is displayed in the inset. (d) Polarization map for $TE_1$ band at α=45° and β=0°. (e) Effects of $\varepsilon_{yy}$ and $\varepsilon_{xy}$ on the variation of BIC positions. (f) The moving trajectory graph of the three BICs as α increases from 0° to 180°.

Next, we take the fixed value of β=0° and change polar angle α to break the in-plane anisotropy symmetry in the PCS. For this case, the off-diagonal term $\varepsilon_{xy}$ is nonzero in the anisotropic material. When we rotate the OA of anisotropic material to α=45° ($\varepsilon_{xy} = 0.1$, $\varepsilon_{yy} = 9.1$), the two accidental BICs shift to $(k_x, k_y)$ =(-0.04, -0.023) 2π/P and (0.04, 0.023) 2π/P along the $k_y$ direction, while the symmetry-protected BIC stays at Γ point, as shown in Fig. 1(d). To carefully investigate the roles of $\varepsilon_{yy}$ and $\varepsilon_{xy}$, in Fig. 1(e) we increase $\varepsilon_{yy}$ from 9.2 to 10 (keeping $\varepsilon_{xy}$ as 0) individually and find that the two accidental BICs will only move slightly in the $k_x$ direction. If we individually move $\varepsilon_{xy}$ from 0 to 0.5 (keeping $\varepsilon_{yy}$ as 9.2), the accidental BICs will move more in the $k_y$ direction. In Fig. 1(f), we finally calculate the evolutionary paths of three BICs as α increases from 0° to 180° in a period. The symmetry-protected BIC is not affected by the $\varepsilon_{xy}$ component, while for the two accidental BICs, they will shift to the farthest position in $k_y$ direction when the $\varepsilon_{xy}$ reaches its maximum magnitude (α=45° and 135°), and then gradually return to the $k_y = 0$ line when $\varepsilon_{xy}$=0 (α=0° and 90°). Therefore, the off-diagonal term $\varepsilon_{xy}$ can simply modulate the positions of accidental BICs in momentum space, and there is no splitting of BICs due to the preservation of out-of-plane symmetry.

In Fig. 2 we rotate OA in the y-z plane and disregard the effect of α on the anisotropic parameters, for this case, the out-of-plane anisotropy symmetry is broken. Taking the angles into Eq. (2), we can see that the off-diagonal term $\varepsilon_{yz}$ is nonzero while $\varepsilon_{xy}$ is zero in the anisotropic material. As we stated above, three BICs ($V_1$, $V_2$ and $V_3$) locate at the $k_y$=0 line if the angles α and β are both 0°, as shown in Fig. 2(a). When β increases from 0° to 10°, the three BICs will split into six C points and move towards negative $k_y$ direction. Fig. 2(b) displays the calculated far-field polarization maps in $k$-space for $β=10°$, where the blue and red ellipses represent LH and RH elliptical polarization, respectively. Clearly, each BIC develops into a pair of C points with the same charge and opposite handedness, i.e., $V_1$ (-1 charge) split into $C_1$ and $C_2$ with -1/2 charge, $V_2$ (+1 charge) split into $C_3$ and $C_4$ with +1/2 charge, $V_3$ (-1 charge) split into $C_5$ and $C_6$ with -1/2 charge. With the increase of $β$, the $C_2$ and $C_3$ points ($C_4$ and $C_5$ points) will move toward each other and annihilate near $β=20°$, leaving only $C_1$ and $C_6$ points, and the two C points will continue to shift until they reach the farthest positions at $(k_x, k_y) =(\pm 0.059, -0.039) 2\pi/P$ for $β=45°$, as displayed in Fig. 2(c). This can be easily understood from Eq. (2) that $\varepsilon_{zy}$ reaches its maximum magnitude of 0.1 at $β=45°$. In Fig. 2(d), we further calculate the positions of different singularities at some specific angles in $k$-space, and illustrate the evolutionary trajectories of these singularities using the directions of arrows in a period of $\tilde{\varepsilon}_A$ ($β=0°\sim 180°$). The process from Fig. 2(a-c) has been represented in the $k_y<0$ region of left panel in Fig. 2(d). When $β$ increases from 45° to 80° in the $k_y<0$ region of right panel, two pairs of C points will generate while keeping the topological charge conserved, as displayed for $β = 80°$. Further increasing $β$ from 80° to 90°, adjacent C points with the same charge and opposite handedness will merge into BICs in this process, forming three singularities with $q=\pm 1$ at $β=90°$ (the off-diagonal term $\varepsilon_{yz}=0$ at this point). For the increase of $β$ from 90° to 180°, similar splitting, annihilation, generation and merging process can be observed in the $k_y>0$ region in momentum space. If we take $α=45°$ in addition to $β=45°$ ($\varepsilon_{xy} = 0.071$, $\varepsilon_{yz}= -0.05$) in Fig. 2(e), the existence of $\varepsilon_{xy}$ will move the $C_1$ and $C_6$ points from two symmetric positions to asymmetric ones in $k$ space. Combined with the effects of $\varepsilon_{yy}$, we confirm that the splitting of BICs here are caused by the off-diagonal term $\varepsilon_{yz}$, which can regulate the transition between BICs and C points and play an important role in the evolution of polarization singularities.

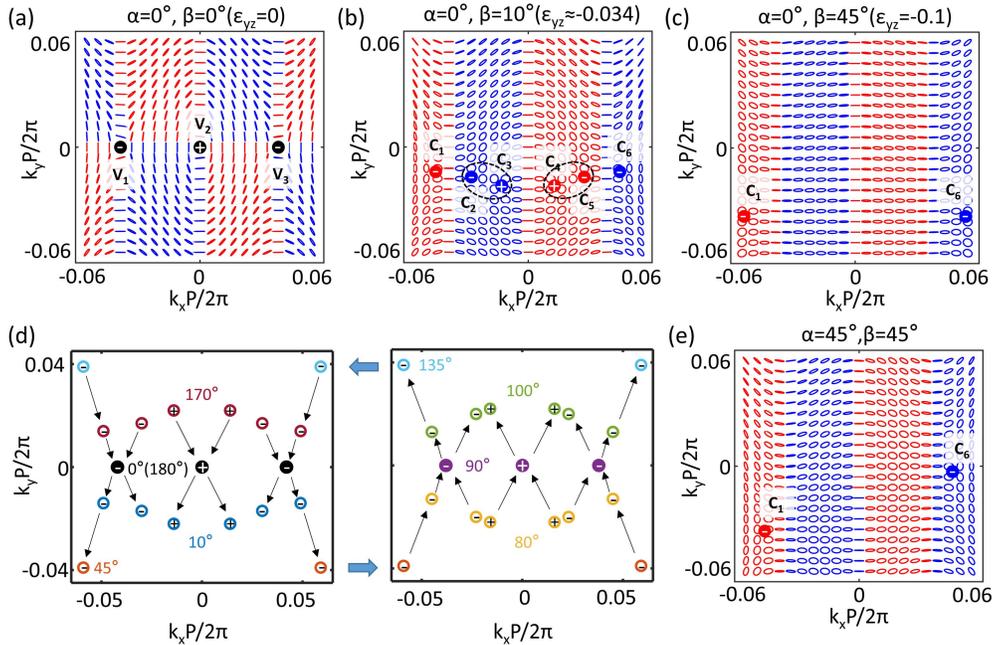

Fig. 2 (a-c) Calculated polarization maps in the momentum space for TE$_1$ band at $β=0°$, $β=10°$, and $β=45°$, respectively, where $α$ is fixed at 0° unchanged. The black solid dots denote BICs with $\pm 1$ charge, and the blue and red solid dots represent left-hand (LH) and right-hand (RH) C point, respectively. (d) Evolution trajectories of C points with the increase of $β$ from 0° to 180°, and $α$ is fixed at 0°. The arrows indicate the direction of movement for C points in a period. (e) The polarization maps for TE$_1$ band at $α=45°$, $β=45°$.

In the above results, we have investigated the evolutions of BICs and multiple C points caused by off-diagonal term $\varepsilon_{yz}$ of the anisotropic material. However, the $C_1$ and $C_6$ points will always evolve towards the trend of three BICs while keeping the topological charge conserved. Here we achieve another evolutionary process by adjusting the parameters of the anisotropic grating lattice in Fig. 1. In detail, we change the height $d=1.09P$ and the initial permittivity tensor of material A as $\tilde{\varepsilon}=\mathrm{diag}(8, 9, 8)$, while keeping other parameters unchanged. By this way, the maximum magnitude of the periodic permittivity components is increased, which may broaden the movement range of polarization singularities in $k$-pace. Fig. 3 shows the evolution of BICs and C points with fixed value of $\alpha=0°$ for example. Staring from the case of $\beta=0°$ in Fig. 3(a), there are three BICs located at $(k_x, k_y)=(0, 0)2\pi/P$ and $(\pm 0.027, 0)2\pi/P$ in the momentum space, respectively. When we slightly increase $\beta$ to $0.6°$, the three BICs split into six C points with the same charge and opposite handedness along $k_x$ direction, and the four C points in the center will quickly annihilate near $\beta\approx 2°$ ($\varepsilon_{yz}=0.035$), leaving only $C_1$ and $C_6$ points. The splitting of BICs here is similar to the results of Fig. 2(a-c). Interestingly, if we continue to increase $\beta$ from $2°$ to $89.9°$, numerical calculations show that only two C points ($C_1$ and $C_6$) exist in this whole process, and there is no extra C points generated, as shown in the case for $\beta=89°$ in Fig.3(c). In the range of $2°$ to $89.9°$, the two C points will first shift along the $-k_y$ direction as $k_x$ and $k_y$ get larger, and then gradually return to the center near the smaller wave vectors. At last, when $\beta=90°$, the $C_1$ and $C_6$ points will merge into a BIC with topological charge as $-1$ at $\Gamma$ point, as displayed in Fig. 4(d). Therefore, the topological charge of at-$\Gamma$ BIC is reversed (the charge switches from +1 to -1) due to charge conservation (the global charge remains $-1$). Different from the results in Fig. 2, by parameter adjustment, there is no six C points in the range of $\beta=45°\sim 90°$ and only an isolated BIC of $q=-1$ at $\Gamma$ point when $\beta=90°$ in the $k$ space.

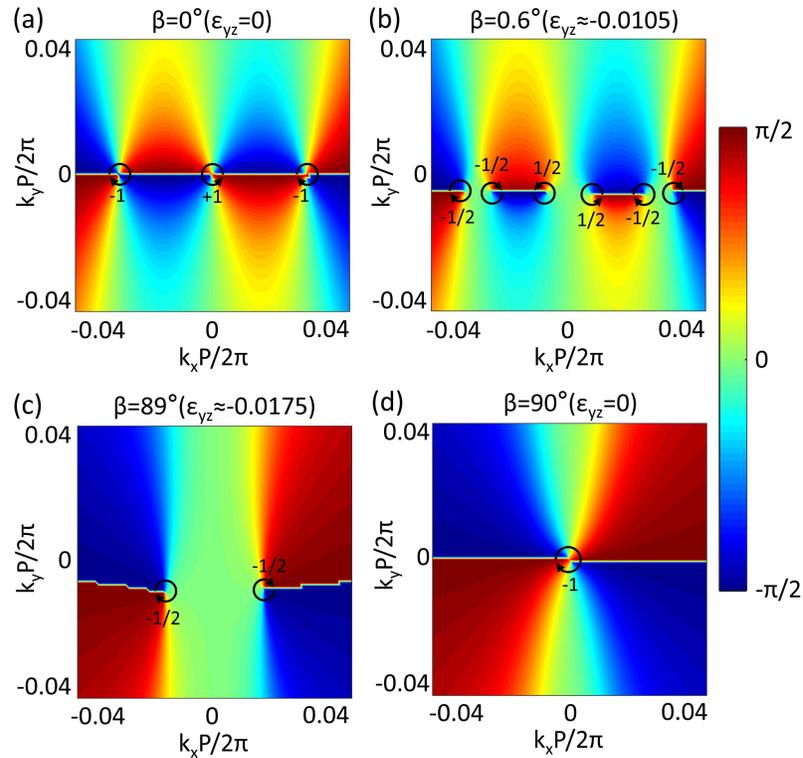

Fig. 3 (a-d) The polarization-orientation-angle diagrams for $TE_1$ band in an anisotropic grating lattice with $d=1.09P$ and the initial permittivity tensor of material A is $\tilde{\varepsilon}=\mathrm{diag}(8, 9, 8)$ for $\beta=0°$ (a), $\beta=0.6°$ (b), $\beta=89°$ (c), $\beta=90°$ (d), respectively, with $\alpha=0°$ unchanged. The C points in (b) from left to right are denoted as $C_1$, $C_2$, $C_3$, $C_4$, $C_5$ and $C_6$, respectively.

*2.2 Evolution behaviors of BICs and C points for TM mode*

In this section, we will investigate the TM mode with $H_y$, $E_x$ and $E_z$ components in the eigensolutions, and investigate the different effects of off-diagonal terms $\varepsilon_{xy}$, $\varepsilon_{xz}$, $\varepsilon_{yz}$ and main-diagonal terms $\varepsilon_{xx}$ and $\varepsilon_{zz}$ on BICs and C points. For the same anisotropic grating system in Fig. 1(a), here we set the grating height $d$ to 1.33P while keep other geometric and material parameters unchanged, and study the evolutions of polarization singularities at different OA orientations for TM mode. Fig. 4(a) shows the TM band structure for the anisotropic grating lattice with $\alpha = 0°$ and $\beta = 0°$, where three BICs with integral topological charge can be observed on the $TM_2$ band, as indicated by the Q factors. It can be found that there are three peaks on the $TM_2$ band, corresponding to the positions of three BICs. Besides, the existence of these BICs is also verified by the ellipticity $\chi$ in Fig. 4(b), where the positions with $\chi=0$ (the polarization is linear) are distributed in a cross shape around each BIC, indicating the V points with ±1 charge. Here, the ellipticity is defined as $\chi=\sin(2\delta)$ with $\tan(\delta)$ being the ratio of polarization ellipse's minor to major axis [40]. When we rotate the OA to $\alpha=45°$ and $\beta=0°$, the two accidental BICs will shift to $(k_x, k_y)=(0.014, -0.01)2\pi/P$ and $(-0.014, 0.01)2\pi/P$, as shown by the ellipticity $\chi$ in Fig. 4(c). Therefore, the effect of $\varepsilon_{xy}$ for TM mode is similar to that in Fig. 1(d) but with different movement directions, and there is no splitting of BICs when the in-plane anisotropy symmetry reduces. Next, we will focus on the three BICs on $TM_2$ band and present the role of $\varepsilon_{yz}$, $\varepsilon_{xz}$, $\varepsilon_{xx}$ and $\varepsilon_{zz}$ on the modulation of polarization singularities.

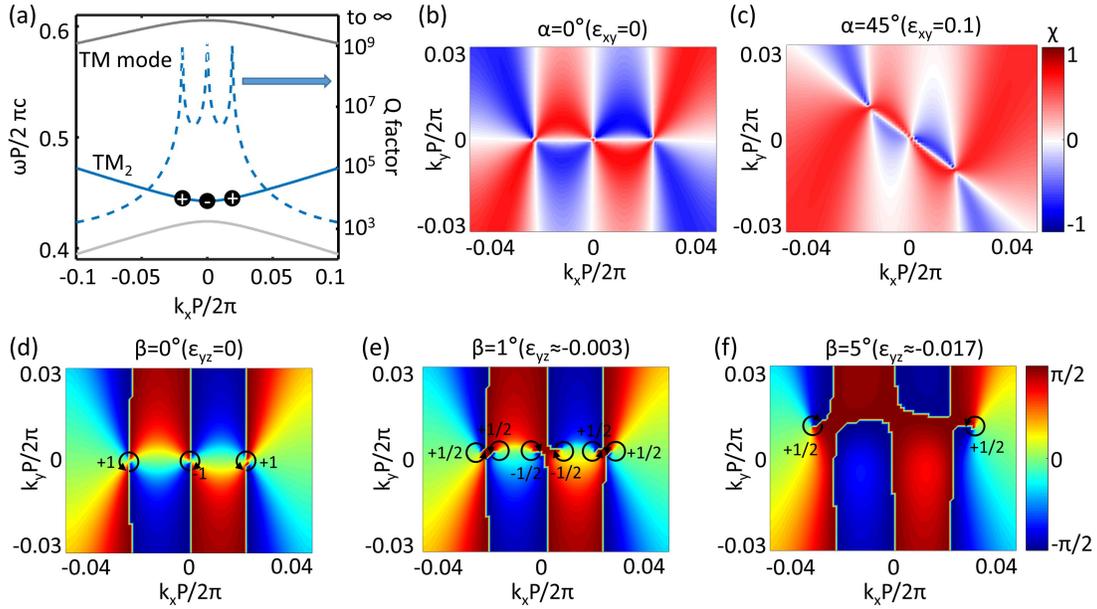

Fig. 4 (a) Simulated TM band structures and Q factors for the anisotropic grating lattice, where the height of the grating lattice is changed to $d=1.33P$. (b, c) The distributions of ellipticity $\chi$ for the $TM_2$ band at $\alpha= 0°$ and $\alpha= 45°$, indicating three V points. (d-f) The polarization-orientation-angle diagrams for $TM_2$ band with $\beta=0°$ (d), $\beta =1°$ (e) and $\beta=5°$ (f), respectively, and the azimuth angle $\alpha$ is set as $0°$. The V points in (d) from left to right are denoted as $V_1$, $V_2$ and $V_3$; the C points in (e) from left to right are denoted as $C_1$, $C_2$, $C_3$, $C_4$, $C_5$ and $C_6$, respectively.

Starting from the case of $\beta = 0°$, we rotate OA in the y-z plane to break the out-of-plane anisotropy symmetry for TM mode, while keeping the polar angle $\alpha =0°$ unchanged, the results are presented in Fig. 4(d-f). For this case, only the off-diagonal term $\varepsilon_{yz}$ is nonzero in the anisotropic material. The polarization orientation diagram in Fig. 4(d) shows that there are three BICs ($V_1$, $V_2$ and $V_3$) locate at the $k_y=0$ line when $\beta = 0°$, with their charges as +1 at $(k_x, k_y) =(\pm0.019, 0) 2\pi/P$ and -1 at $(k_x, k_y) =(0, 0)$, respectively. Different from that in TE mode, the zero phase of polarization for TM mode is on the $k_y=0$ line. When $\beta$ slightly increases from $0°$ to $1°$, the three BICs will split into six C points along the $k_x$ direction and move towards the positive $k_y$ direction. Fig. 4(e) displays the polarization orientation map in $k$-space for $\beta=1°$, where each BIC develops into a pair of C points with the same charge and opposite handedness, i.e., $V_1$ (+1 charge) split into $C_1$ and $C_2$ with +1/2 charge, $V_2$ (-1 charge) split into $C_3$ and $C_4$ with -1/2 charge, $V_3$ (+1 charge) split into $C_5$ and $C_6$ with +1/2 charge. If we further increase $\beta$ to 5°, the $C_2$ and $C_3$ points ($C_4$ and $C_5$ points) will annihilate each other, leaving only $C_1$ and $C_6$ points at $(k_x, k_y) =(\pm 0.027, 0.01) 2\pi/P$, as displayed in Fig. 4(f). Therefore, the off-diagonal term $\varepsilon_{yz}$ also causes the splitting of BICs for the TM mode, though the splitting process here is similar to that in Fig. 2, the TM mode has a completely different polarization distribution from the TE mode. If we continue to change $\beta$ in a period, the generation and merging of C points will also happen as that in TE mode, thus we will not repeat it here.

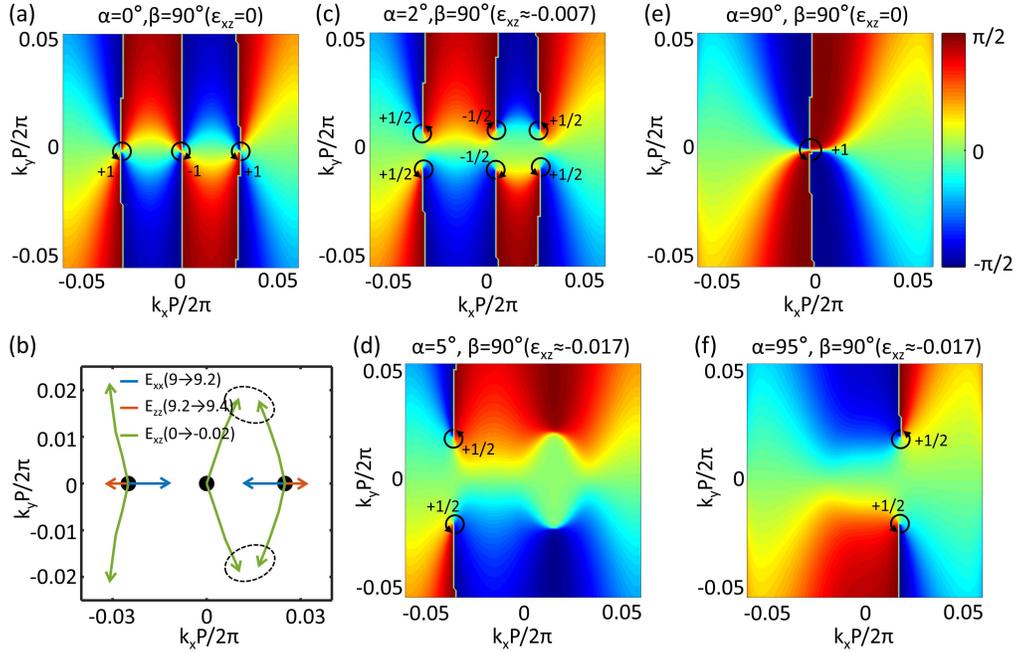

Fig. 5 The polarization-orientation-angle diagrams for $TM_2$ band with α=0° (a), α=2° (c) , α=5° (d), α=90° (e) and α=95° (f), respectively, where β is fixed at 90° unchanged. (b) Effects of $\varepsilon_{xx}$, $\varepsilon_{zz}$ and $\varepsilon_{xz}$ on the variation of BIC and C positions. Blue lines ($\varepsilon_{zz}$=9.2, $\varepsilon_{xz}$=0); Orange lines ($\varepsilon_{xx}$=9, $\varepsilon_{xz}$=0); Green lines ($\varepsilon_{zz}$=9.2, $\varepsilon_{xx}$=9). The V points in (a) from left to right are denoted as $V_1$, $V_2$ and $V_3$, respectively.

Finally, we aim to investigate the effects of $\varepsilon_{xz}$, $\varepsilon_{xx}$ and $\varepsilon_{zz}$ on these polarization singularities, which are unique to the TM mode and make a big difference. Considering the anisotropic permittivity tensor in Eq. (2), we take the fixed value of $\beta$ =90° and change the azimuth angle α, the permittivity tensor is expressed as $(\varepsilon_1\cos(\alpha)^2 + \varepsilon_2\sin(\alpha)^2 \; 0 \; -\Delta\sin(\alpha)\cos(\alpha); \; 0 \; \varepsilon_1 \; 0; \; -\Delta\sin(\alpha)\cos(\alpha) \; 0 \; \varepsilon_1\sin(\alpha)^2 + \varepsilon_2\cos(\alpha)^2$, where $\varepsilon_{xy}$ and $\varepsilon_{yz}$ are both zero. The simulated results are displayed in Fig. 5. Starting from the case of α=0°, $\beta$ =90° in Fig. 5(a), three BICs locate at $V_1$: $(k_x, k_y)$ =(-0.025, 0) 2π/P, $V_2$: $(k_x, k_y)$ =(0, 0) and $V_3$: $(k_x, k_y)$ =(0.025, 0) 2π/P, respectively. Next, we independently control each component of $\varepsilon_{xz}$, $\varepsilon_{xx}$ and $\varepsilon_{zz}$ to study their effects, as shown in Fig. 5(b). Increasing $\varepsilon_{xx}$ ($\varepsilon_{zz}$) will cause the two accidental BICs to move closer to (away from) Γ point, while nonzero $\varepsilon_{xz}$ will result in the splitting of each BIC along $k_y$ direction. After discussing the role of each component, we gradually rotate OA to break the anisotropy symmetry in x-z plane, the results in displayed in Fig. 5(c-f). For α = 2°, the three BICs will split into pairs of C points along $k_y$ direction, forming $C_1$ and $C_2$ points at (-0.028, 0.007) 2π/P and (-0.028, -0.007) 2π/P, $C_3$ and $C_4$ points at (0.004, 0.008) 2π/P and (0.004, -0.008) 2π/P, $C_5$ and $C_6$ points at (0.022, 0.008) 2π/P and (0.022, -0.008) 2π/P, respectively, as the polarization-orientation map shown in Fig. 5(c). Besides, the three C points in upper row are the LH states, while those in the bottom row are the RH states. Continuing to increase α to 5° (see Fig. 5(d)), the $C_3$ and $C_5$ points ($C_4$ and $C_6$ points) with the opposite charges and same handedness will annihilate each other, leaving only $C_1$ and $C_2$ points at (-0.03, ±0.018) 2π/P. In contrast to previous results, here the remaining two C points at α =5° come from one BIC ($V_1$). Moreover, if we continue to increase α from 5° to 89.9°, it is found that there will always be two C points with their positions shifting in *k*-space, that is, these two C points will first move towards the larger $k_y$ direction and then gradually move towards the $k_y$=0 direction, keeping approaching the $k_x$=0 direction simultaneously. When α = 90°, they will eventually merge into a BIC at Γ point with charge q=+1, indicating the reversal of topological charge for Γ-BIC, as shown in Fig. 5(e). On the other hand, when α increases to 95°, the Γ-BIC will split into a pair of C points and move to the $k_x$>0 range, with their charges as 1/2. It is foreseeable that another two pairs of C points will appear in $k_x$<0 range and these six C points will then finally merge into three BICs as α changes from 95° to 180°. Overall, the results in Fig. 5 elaborate that a nonzero $\varepsilon_{xz}$ can cause the BICs in TM mode to split along $k_y$ direction and enrich the polarization distributions of far-field radiation, which is quite interesting and different from the results for TE mode.

## 3. Conclusions

In summary, we have realized diverse evolutions of multiple BICs and C points in an anisotropic grating lattice and revealed the effects of different off-diagonal terms on polarization singularities for both TE and TM modes, respectively. By rotating OA in the x-y plane, accidental BICs can only be shifted without splitting, while rotating OA out of the x-y plane, the three BICs are split into six C points and accompanied by the annihilation, generation and merging of C points. Additionally, we also show that the evolutionary process of these polarization singularities can be tuned by changing the off-diagonal permittivity tensor, where the charge reversal at Γ point can be realized. This extra degree of freedom offers a flexible way to modulate topological singularities, and may have potential applications in the fields of chiral light-field manipulation.

**Acknowledgements.** This work was supported by the National Natural Science Foundation of China (12104169, 12361161667), Natural Science Foundation of Fujian Province (2024J01081), and the Fundamental Research Funds for the Central Universities (ZQN-1110).

**Disclosures.** The authors declare no conflict of interests.

**Data availability.** Data underlying the results presented in this paper are not publicly available at this time but may be obtained from the authors upon reasonable request.

**APPENDIX A: DERIVATION OF THE ANISOTROPIC PERMITTIVITY TENSOR**

The permittivity tensor of anisotropic material A for an arbitrary OA orientation can be calculated from coordinate transformation using rotation matrix R(α) and R(β). We take the initial relative permittivity tensor of material A as $\tilde{\varepsilon}$=diag($\varepsilon_1$, $\varepsilon_2$, $\varepsilon_1$) when the OA is oriented along x direction. For the first step, we rotate in the x-y plane using rotation matrix R(α) = [cos(α) sin(α) 0; -sin(α) cos(α) 0; 0 0 1], which can be written as

$$\tilde{\varepsilon}' = R(\alpha)\tilde{\varepsilon}R(\alpha)^T = \begin{bmatrix} \varepsilon_1\cos(\alpha)^2 + \varepsilon_2\sin(\alpha)^2 & (\varepsilon_2 - \varepsilon_1)\sin(\alpha)\cos(\alpha) & 0 \\ (\varepsilon_2 - \varepsilon_1)\sin(\alpha)\cos(\alpha) & \varepsilon_1\sin(\alpha)^2 + \varepsilon_2\cos(\alpha)^2 & 0 \\ 0 & 0 & \varepsilon_1 \end{bmatrix}, \quad (A1)$$

Next, we rotate in the y-z plane using rotation matrix R(β) = [1 0 0; 0 cos(β) sin(β); 0 -sin(β) cos(β)], the permittivity tensor can be written as:

$$\tilde{\varepsilon}_A = R(\beta)\tilde{\varepsilon}'R(\beta)^T =$$
$$\begin{bmatrix} \varepsilon_1\cos(\alpha)^2 + \varepsilon_2\sin(\alpha)^2 & \Delta\sin(\alpha)\cos(\alpha)\cos(\beta) & -\Delta\sin(\alpha)\cos(\alpha)\sin(\beta) \\ \Delta\sin(\alpha)\cos(\alpha)\cos(\beta) & \Delta\cos(\alpha)^2\cos(\beta)^2 + \varepsilon_1 & -\Delta\sin(\beta)\cos(\beta)\cos(\alpha)^2 \\ -\Delta\sin(\alpha)\cos(\alpha)\sin(\beta) & -\Delta\sin(\beta)\cos(\beta)\cos(\alpha)^2 & \Delta\cos(\alpha)^2\sin(\beta)^2 + \varepsilon_1 \end{bmatrix}, \quad (A2)$$

where $\Delta = \varepsilon_2 - \varepsilon_1$. Through this operation, the material's permittivity for different polar angle $\alpha$ and azimuth angle $\beta$ can be obtained.

**APPENDIX B: WAVE EQUATIONS IN THE ANISOTROPIC MEDIUM**

The Maxwell's equations in electric anisotropic material A with relative permittivity $\tilde{\varepsilon}_A$ and relative permeability $\mu_r=1$ are expressed as

$$\nabla \times E = i\omega\mu_0 H$$
$$\nabla \times H = -i\omega\varepsilon_0 \tilde{\varepsilon}_A E' \tag{A3}$$

and the wave equation can be obtained as

$$\nabla^2 E - \nabla(\nabla \cdot E) + \omega^2 \mu_0 \varepsilon_0 \tilde{\varepsilon}_A E = 0, \tag{A4}$$

Considering the plane waves with space factor $e^{ik_x x + ik_y y + ik_z z}$, the wave equation for E becomes

$$k^2 E - k(k \cdot E) - \omega^2 \mu_0 \varepsilon_0 \tilde{\varepsilon}_A E = 0, \tag{A5}$$

$$k^2 \begin{bmatrix} E_x \\ E_y \\ E_z \end{bmatrix} - (k_x E_x + k_y E_y + k_z E_z) \begin{bmatrix} k_x \\ k_y \\ k_z \end{bmatrix} - \omega^2 \mu_0 \varepsilon_0 \begin{bmatrix} \varepsilon_{xx} & \varepsilon_{xy} & \varepsilon_{xz} \\ \varepsilon_{yx} & \varepsilon_{yy} & \varepsilon_{yz} \\ \varepsilon_{zx} & \varepsilon_{zy} & \varepsilon_{zz} \end{bmatrix} \begin{bmatrix} E_x \\ E_y \\ E_z \end{bmatrix} = 0, \tag{A6}$$

Then the Eq. (A4) can be simplified as

$$\begin{bmatrix} \omega^2 \mu_0 \varepsilon_0 \varepsilon_{xx} - k_y^2 - k_z^2 & \omega^2 \mu_0 \varepsilon_0 \varepsilon_{xy} + k_x k_y & \omega^2 \mu_0 \varepsilon_0 \varepsilon_{xz} + k_x k_z \\ \omega^2 \mu_0 \varepsilon_0 \varepsilon_{yx} + k_x k_y & \omega^2 \mu_0 \varepsilon_0 \varepsilon_{yy} - k_x^2 - k_z^2 & \omega^2 \mu_0 \varepsilon_0 \varepsilon_{yz} + k_y k_z \\ \omega^2 \mu_0 \varepsilon_0 \varepsilon_{zx} + k_x k_z & \omega^2 \mu_0 \varepsilon_0 \varepsilon_{zy} + k_y k_z & \omega^2 \mu_0 \varepsilon_0 \varepsilon_{zz} - k_x^2 - k_y^2 \end{bmatrix} \begin{bmatrix} E_x \\ E_y \\ E_z \end{bmatrix} = 0. \tag{A7}$$

where the electric field $\mathbf{E} = (E_x, E_y, E_z)e^{ik_x x + ik_y y + ik_z z}e^{-i\omega t}$ and $k$ is the wavevector. Two eigensolutions can be solved from Eq. (A5) for TE and TM modes, respectively. For TE mode with $E_y \neq 0$, $E_x = E_z = 0$, we can obtain

$$\begin{aligned} k_x k_y + \omega^2 \mu_0 \varepsilon_0 \varepsilon_{xy} &= 0, \\ k_y k_z + \omega^2 \mu_0 \varepsilon_0 \varepsilon_{zy} &= 0, \\ k_x^2 + k_z^2 &= \omega^2 \mu_0 \varepsilon_0 \varepsilon_{yy}, \end{aligned} \tag{A8}$$

For TM mode with $E_y=0$, $E_x \neq 0$, $E_z \neq 0$, the eigensolution can be solved as

$$\frac{E_x}{E_z} = \frac{k_x k_z + \omega^2 \mu_0 \varepsilon_0 \varepsilon_{xz}}{k_y^2 + k_z^2 - \omega^2 \mu_0 \varepsilon_0 \varepsilon_{xx}} = \frac{k_y k_z + \omega^2 \mu_0 \varepsilon_0 \varepsilon_{yz}}{-(k_x k_y + \omega^2 \mu_0 \varepsilon_0 \varepsilon_{yx})} = \frac{k_x^2 + k_y^2 - \omega^2 \mu_0 \varepsilon_0 \varepsilon_{zz}}{k_x k_z + \omega^2 \mu_0 \varepsilon_0 \varepsilon_{zx}}. \tag{A9}$$

Therefore, for an axisymmetric anisotropic material (along the principal axis), we can find that for TE mode, the eigensolution are dependent of $\varepsilon_{yy}$, $\varepsilon_{xy}$, $\varepsilon_{yz}$. However, for TM mode, the eigensolution are related to all three off-diagonal terms $\varepsilon_{xy}$, $\varepsilon_{xz}$ and $\varepsilon_{yz}$ and two main-diagonal terms $\varepsilon_{xx}$ and $\varepsilon_{zz}$.